\documentclass[acmsmall]{acmart}
\usepackage{subfiles}
\usepackage{graphicx}
\usepackage{caption}
\usepackage{subcaption}
\usepackage{multirow}
\usepackage{xspace}
\usepackage{algorithm}
\usepackage{algorithmic}
\usepackage{amsmath}
\usepackage{mathtools}
\usepackage[skip=2pt]{caption}
\usepackage[skip=2pt]{subcaption}
\usepackage{balance}  
\usepackage{hyperref}
\usepackage{amsmath}
\usepackage[export]{adjustbox}
\usepackage{tikz}

\usepackage{makecell}

\newcommand{\out}[1]{{#1}}

\newcommand{\new}[1]{\out{{}}}
\newcommand{\news}[1]{\out{{}}}

\AtBeginDocument{%
  \providecommand\BibTeX{{%
    \normalfont B\kern-0.5em{\scshape i\kern-0.25em b}\kern-0.8em\TeX}}}

\setcopyright{acmlicensed}
\copyrightyear{2020}
\acmYear{2020}
\acmDOI{TBD}

\acmConference[CSCW '20]{CSCW '20: ACM Conference on Computer-Supported Cooperative Work and Social Computing}{2020}{}
\acmPrice{15.00}
\acmISBN{TBD}

\begin{document}

\title[Programming Education via Live Streaming]{Towards Supporting Programming Education at Scale via Live Streaming}


\author{Yan Chen}
\affiliation{%
 \institution{University of Michigan}
 \city{Ann Arbor}
 \state{MI}
 \country{USA}}
\email{yanchenm@umich.edu}
\authornote{This work was completed while the author was an intern at Google.}

\author{Walter S. Lasecki}
\affiliation{%
 \institution{University of Michigan}
 \city{Ann Arbor}
 \state{MI}
 \country{USA}}
\email{wlasecki@umich.edu}

\author{Tao Dong}
\affiliation{%
 \institution{Google Inc.}
 \city{Mountain View}
 \state{CA}
 \country{USA}}
\email{taodong@google.com}

\renewcommand{\shortauthors}{Chen et al.}

\begin{abstract}
  
Live streaming, which allows streamers to broadcast their work to live viewers, is an emerging practice for teaching and learning computer programming. Participation in live streaming is growing rapidly, despite several apparent challenges, such as a general lack of training in pedagogy among streamers and scarce signals about a stream's characteristics (e.g., difficulty, style, and usefulness) to help viewers decide what to watch. 
To understand why people choose to participate in live streaming for teaching or learning programming, and how they cope with both apparent and non-obvious challenges, we interviewed 14 streamers and \new{12} viewers about their experience with live streaming programming.
Among other results, we found that the casual and impromptu nature of live streaming makes it easier to prepare than pre-recorded videos, and viewers have the opportunity to shape the content and learning experience via real-time communication with both the streamer and each other.
Nonetheless, we identified several challenges that limit the potential of live streaming as a learning medium. For example, streamers voiced privacy and harassment concerns, and existing streaming platforms do not adequately support viewer-streamer interactions, adaptive learning, and discovery and selection of streaming content.
Based on these findings, we suggest specialized tools to facilitate knowledge sharing among people teaching and learning computer programming online, and we offer design recommendations that promote a healthy, safe, and engaging learning environment.

\end{abstract}



\maketitle

\section{Introduction}

In the last few years, video-centric learning materials from online platforms such as YouTube, Coursera, and Khan Academy have enabled people of all ages and countries to share and gain knowledge~\cite{youtube,coursera,khanacademy}. 
These videos are often well-structured, pre-recorded, and carefully edited by experienced instructors or even professional teams, and the content often covers specific topics, concepts, or techniques. 
However, the effort required to start sharing knowledge using these approaches is often high. 
Moreover, the asynchronous interaction and non-immediate feedback can impede learning and reduce student engagement~\cite{branon2001synchronous, wilson2011success}.

Live streaming has increasingly become a way to share and obtain technical knowledge. 
To understand this emerging practice, we studied live video streaming of computer programming where the focus is on demonstrating knowledge and practical skills for educational purposes.
We investigate this phenomenon for two reasons: 1) there is widespread interest in learning to program~\cite{moocs}, and 2) prior work has found that live coding, in which instructors write code from scratch on a computer connected to a projector in a lecture~\cite{paxton2002live}, is effective for programming education~\cite{rubin2013effectiveness}. Live streaming platforms scale and democratize this pedagogical technique for anyone who has a decent Internet connection and a passion for sharing their knowledge with the world. 

Although live streaming is a promising format in the online learning ecosystem~\cite{faas2018watch}, there are apparent downsides. Compared to massive open online courses (MOOCs) or other professional educational settings, streamers often lack pedagogical training, leading to large variations in content quality, presentation skills, and the ability to engage students. 
For viewers, there are few signals to determine whether an upcoming stream will be a good fit for the viewer's learning needs and level of expertise, because unlike pre-recorded videos, live streams have no ratings, view counts, summaries, or previews. 
\new{Our interviews also revealed concerns about harassment or spam in the chat from both viewers and streamers, though these issues were not as common as they are in other streaming content.~\cite{seering2017shaping}}
These challenges may negatively affect the viewer's learning experience, making live streaming a less effective learning medium compared to pre-recorded videos.


To sustain and improve the developing practice of live streaming programming, we aim to understand the reasons for the disconnect between the growing interest in  stream participation and the aforementioned challenges.
In particular, we explore the benefits that the major stream platforms (e.g., YouTube Live, Twitch) provide for learning-oriented live streams, and the challenges people overcome, work around, or fail to tackle on those platforms. 
We use pre-recorded video, a more established medium for learning programming, as a reference point to gain a better understanding of the position of live streaming in the online programming education ecosystem. Specifically, we wanted to answer the following research questions. 
\\

\vspace{-0.8pc}
\textbf{RQ1:} why do streamers choose to share programming knowledge via live streaming?

\textbf{RQ2:} why do viewers choose to learn programming via live streaming?

\textbf{RQ3:} what barriers do streamers face when sharing programming knowledge via live streaming?

\textbf{RQ4:} what barriers do viewers face when learning programming via live streaming?
\\
\vspace{-0.5pc}

Through 26 in-depth interviews (14 streamers, 12 viewers), we found that 1) the nature of live (e.g., more casual) content makes streaming sessions easier to prepare than pre-recorded video; 2) streaming gives viewers the opportunity to shape the content through real-time interactions; 3) live interactions make these streams infeasible to script, intensifying streamers' privacy and harassment concerns; and 4) inadequate tool support makes learning hard to personalize for individual viewers (e.g., pacing). 

Based on these findings, we discuss four implications for the design of streaming platforms, including a) evolving generic video streaming platforms to support personalized learning, b) reducing the effort of documenting and retrieving points of interest in live streams, c) improving support for content moderation, and d) mitigating streamers' privacy and security risks. Also, we critique and elaborate on prior design prototypes and concepts applicable to this domain. These ideas work toward a form of improvisational, real-time, and personalized knowledge sharing to fill gaps in formal programming education. In sum, we contribute the following:

\begin{itemize}
    \vspace{-0.2pc}
    \item We position live streaming in the online, and especially video-based, programming education ecosystem by highlighting unique advantages that can be leveraged, and limitations that must be mitigated and managed to make this learning medium more effective and viable. 
    \vspace{0.1pc}
    \item We draw design implications to facilitate programming knowledge sharing via live streaming, as well as design modifications to promote a healthier and more secure learning environment.
\end{itemize}

\section{Related Work}
Our work mainly relates to two fields: common practices of programming learning and live streaming. 
In this section, we will review the key work in these two areas. 

\subsection{Common Practices for Learning Computer Programming}
\subsubsection{Informal Learning}

Consuming knowledge on streaming platforms like Twitch.tv is a form of informal learning, which is a type of learning undertaken on the learner's own, without externally imposed criteria~\cite{selwyn2007web}. 
Unlike formal learning, which is often institutionally sponsored and structured, informal learning provides learners better control and personalization, representing a promising opportunity for learners to develop different kinds of skills through lifelong learning~\cite{bransford2007preparing}.
In recent years, a growing number of people have begun to engage in a wide range of technology-based informal learning opportunities.
However, the Educating the Engineers of 2020 report highlights the lack of research on informal learning within engineering education~\cite{phase2005educating}.


In the context of live streaming, prior work has reported on the unique learning relationship that this medium creates for programming communities.
More specifically, the act of learning via live streaming often takes place in a collaborative manner rather than an individual one, which enables a sense of social presence. 
However, researchers have raised the concern that the place of live streaming within the online learning ecosystem is still unclear, making it difficult for people to choose when to use this medium and what benefits they might receive from it~\cite{faas2018watch}. 
Our work aims to more clearly position live streaming in the ecosystem of different learning mediums, adding new understanding to the body of literature on informal learning within programming education~\cite{phase2005educating}.

\subsubsection{Video Lecturing}

The use of video lecture capture in higher education is becoming increasingly commonplace in universities worldwide and massive open online courses (MOOCs)~\cite{holliman2013mediating,vardi2012will}.
Video lectures provide extra resources that may complement students' studies and add flexibility, temporally and spatially, for students to catch up and revisit the learning materials~\cite{kruger05diss}. 
Prior work has explored the cognitive value of educational videos on YouTube and found a set of cognitive features that correlates with students' learning gains~\cite{shoufan2019estimating}.
However, video lectures have a negative impact on students' performance compared to regular classroom lectures. 
Prior work has shown that video lectures can be perceived as uninteresting and may lack beneficial student-teacher and student-student interactions, leading to low class attendance and less engagement~\cite{kim2014understanding, branon2001synchronous, wilson2011success}.

Video delivery of programming solutions may be useful in enabling a lecturer to illustrate the complex decision-making processes and the incremental nature of the actual code development process. 
McGowan et al. analyzed students' activities during programming lecture videos and found that code-based, practical demonstrations are more valued by students than solo explanations~\cite{mcgowan2016teaching}.
Our work explores the potential pedagogical value of coding demonstration via live streaming to inform and improve its educational benefits. We focus specifically on the informal learning context rather than institutionally structured alternatives such as the types of content found in MOOCs.



\subsubsection{Live Coding}

The practice that our work focuses on closely relates to live coding, an existing common pedagogical practice for teaching computer programming where instructors write code with little preparation work and project the process to learners~\cite{paxton2002live}.
Learners benefit from watching instructors' iterative process of thinking, designing, coding, and testing, which is rarely covered in slides-driven lectures~\cite{gantenbein1989programming, bennedsen2005revealing}.
Prior work has explored the effectiveness of live coding for teaching introductory programming material and found it to be on par with using static code examples~\cite{rubin2013effectiveness}.
Raj et al.~\cite{raj2018role} found that students prefer live-coding, as they benefit from seeing the process of programming and debugging.  
Our work will explore the benefits of the live coding practice in online informal learning settings.

\subsection{Live Streaming}

Recently, an increasing amount of work has focused on studying live streaming in the human-computer interaction (HCI) community,
as this medium enables anyone to share their life experiences, opinions, and knowledge on a variety of topics. 
Prior work has explored the use of streaming platforms and found that streams involve a diverse range of activities, with self-promotion as the most common motivation~\cite{tang2016meerkat}.
Dougherty explored the use of Qik, showing that 11\% of videos had civic importance, such as journalistic and activist value~\cite{dougherty2011live}. 
Researchers have also explored the social norms of streaming given its ``live'' quality. 
Seering et al. found that viewers could shape the streaming content, and have an impact on other viewers' behavior through interactions such as moderation interventions~\cite{seering2017shaping}.
In this section, we will first discuss prior work on entertainment-focused live streaming, which is the most popular type of live content on our target streaming platforms (i.e., Twitch.tv, YouTube Live). Then we will review work on knowledge-sharing-focused and programming-focused live streaming and discuss their differences from entertainment-focused live streaming.

\subsubsection{Entertainment-focused Live Streaming}
Live streaming has been primarily focused on entertainment content, such as eSports~\cite{hamari2017esports,pellicone2017game}, live events~\cite{haimson2017makes, tang2017crowdcasting}, outdoor activities~\cite{lu2019vicariously}, or news updates~\cite{fbnews,trump}. Streaming platforms like Twitch, which was founded in 2012, originally focused on game-related content. 
Prior work in entertainment-focused live streaming found that viewers engage in live streaming because of the unique content of the stream, and being able to participate in that stream's community~\cite{hamilton2014streaming}. 
They also revealed a series of issues in the live streaming community, namely vicious trolling and harassment~\cite{seering2017shaping,wohn2019volunteer}.

\subsubsection{Knowledge-sharing-focused Live Streaming}
Other work has also studied knowledge sharing via live streaming. 
Fraser et al. explored how people share knowledge via creative live streams where the content focuses on creating a novel artifact, such as digital visual art~\cite{fraser2019sharing}. 
Chen et al. studied multimedia tools to enrich interactions in live streams for language learning ~\cite{samat2019live}.
Lu et al. investigated the sharing of intangible cultural heritage artifacts (ICH) in China via live streaming~\cite{lu2019feel}. 
They found that streamers' motivations were less financially driven and more altruistic: streamers are driven to develop online communities with unique identities and to share knowledge through meaningful conversations. 
Other than skill-related knowledge-sharing, Lu et al. found 68\% of their survey respondents use live streaming to share their life experiences, such as dealing with pressure from work~\cite{lu2018you}.
They found that sharing life experiences in the stream encourages more conversational interactions between streamers and viewers, giving viewers opportunities to learn more about streamers and making them more willing to continue following the streamers.

Unlike entertainment-focused live streaming, streamers in knowledge-sharing streams found it challenging to keep their viewers engaged while focusing on more performative activities~\cite{fraser2019sharing}. 
Similarly, viewers found these knowledge-sharing streams to be less efficient and less engaging in terms of streamer-viewer interaction than entertainment-focused streams~\cite{yang2020snapstream}. 
Also, viewers wished the stream archives had better documentation for further consumption~\cite{lu2018streamwiki}.
These concerns imply that existing streaming platforms need to make changes to adapt to the interaction needs in live streams focused on knowledge-sharing.


\subsubsection{Programming-focused Live Streaming}
Growing interest in the online streaming community has led to increased research related to programming-focused live streams as a phenomenon in CS education. 
For example, Haaranen identified educational moments in the chats during programming streams~\cite{haaranen2017programming}. 
Faas et al. conducted a participant-observer study that observed nine streamers and interviewed five streamers and four viewers regarding their live streaming programming experiences~\cite{faas2018watch}. 
The authors reported that viewers played a key role while watching the stream due to their interactions with the streamer and each other. 
They also found a mutual mentorship among viewers and streamers both during and outside of the stream.

Alaboudi et al. conducted a similar preliminary study that compared live streaming programming to pair programming by watching 20-hour stream archive videos and surveying five streamers~\cite{alaboudi2019exploratory}. 
They found that streamers struggled to balance between engaging with viewers and streaming the content, and viewers who joined the stream in the middle found it difficult to understand the stream content.


All these studies have suggested that programming-focused live streams have great potential as a unique form of online learning, yet these streams face a few unique challenges compared to those of streamers that focus on other content. Those challenges were under-examined in prior work.
First, unlike game playing or visual art creation, programming often involves frequent application switching, from the integrated development environment (IDE) and the program output to browsers and other support tools~\cite{brandt2009two}. 
This application switching behavior makes it more difficult to follow a programming-focused live stream, as streamers often share one application at a time, so if viewers do not follow along, they might miss the context and thus fall behind. 
Additionally, debugging and on-the-fly learning (for streamers) are common activities while developing software. 
However, these activities can be time consuming and tedious, making the stream less engaging. 
Finally, software development often requires a series of configuration settings, meaning streamers must provide personal information (e.g., API keys, passwords), which could cause privacy issues. 

\section{Methods}

To better understand these challenges and fill in the research gaps, we conducted semi-structured interviews with 26 participants who engage in live streaming programming. 
We recruited our participants via purposive sampling combined with snowball sampling.
The first eight streamers were recruited based on their high ranking in Twitch.tv's programming and software development directories and in YouTube's search results with queries such as ``live streaming programming'' and ``live programming.'' 
The rest of the streamers and all of the viewers were recruited through an institutionally-maintained general participant pool and snowball sampling.
We reached out to interview candidates by email and Twitter with a short description of our research purpose and a background survey about their live streaming experience and demographic information.
Based on their responses, we selected our final interviewees (refers to both streamers and viewers) to ensure the relevance of their experiences, as well as consider diversity in gender, age, and occupation. 



Each interview session was conducted via video conferencing software and lasted 40 to 75 minutes.
Following their interview, each participant received a thank you gift valued appropriately for their time (above standard market rate for programmers in the U.S.).
We designed the interview based on our research questions mentioned in the introduction.
We began by asking interviewees about their general motivations for producing or watching live streams compared to pre-recorded videos. 
We asked \news{streamers} when and how they started live streaming, why they continue participating, how they grew their community, and their reasons for streaming more or less often. 
\news{We asked viewers when and how they started watching live streams, how often they watch, and their reasons for watching more or less often. }
We then asked interviewees to open an archived stream they recently hosted or watched. 
We sought to understand what kinds of interactions interviewees had with others before, during, and after live streaming sessions and what their intentions and concerns were. 
We asked interviewees about their best practices for protecting their or others' sensitive information, for information about their concerns regarding live streaming, and what they wished to improve or to see improved regarding their privacy concerns.

We used an automated transcription service to transcribe all the interviews.
The lead author first went through the transcripts along with the audio, corrected machine transcription mistakes, and then exported the transcripts to a Google Document. 
The lead author then conducted iterative coding to divide the interview responses into major themes using an inductive analysis approach~\cite{corbin2015basics}.
The lead author coded the first few interviews to form an initial codebook, and then discussed with other authors to merge similar codes and identify important emerging themes. 
With the new codebook, the lead author coded all 26 interviews while discussing and iterating key themes with other authors from time to time.
We lightly edited interviewees' quotes for readability.


\begin{table}
 \centering
 \begin{tabular}{ p{1cm} |p{1.8cm}  | p{2cm} | p{1.5cm} | p{1.8cm}| p{2.5cm} }
  \textbf{PID} & \textbf{Occupation} & \textbf{\new{Gender}} & \textbf{Age}  & \textbf{Platform}& \textbf{Programming Experience (yr)} \\
  \midrule
   S1 & P & M  & 18 to 23 & Y & 15\\
   \hline
   S2 & P & M & 41 to 50 & Y,T  & 4\\
   \hline   
   S3 & P & M & 18 to 23 & Y & 3\\
   \hline
   S4 & P & M & 31 to 40 & Y & 2.5\\
   \hline
   S5 & P & M & 31 to 40 & Y,T & 20 \\
   \hline
   S6 & P & M & 41 to 50 & Y & 4\\
   \hline
   S7 & H & M & 24 to 30 & Y,T & 15\\
   \hline
   S8 & S & M & 18 to 23 & Y & 6\\
   \hline
   S9 & P & F  & 24 to 30 & Y & 10+\\
   \hline
   S10 & P & Prefer not to say & Prefer not to say & T & 26\\
   \hline
   S11 & P& F & 24 to 30 & T & 10\\
   \hline
   S12 & P & F & 31 to 40 & Y,T & 10\\
   \hline
   S13 & H & F & 24 to 30 & T & 15\\
   \hline
   S14 & P & F & 31 to 40 & T & On and off 20\\
   \hline
   V1 & H  & F & 18 to 23 & Y & 2\\
   \hline
   V2 & S & M & 18 to 23 & Y & 8\\
   \hline   
   V3 & S & M & 24 to 30 & T & 5\\
   \hline
   V4 & P & M & 18 to 23 & T & 3\\
   \hline
   V5 & S & M & 24 to 30 & Y &4\\
   \hline
   V6 & P & M & 41 to 50 & Y &21\\
   \hline
   \new{V7} & P & F & 18 to 23 & Y & 4\\
   \hline
   \new{V8} & P & M & 24 to 30 & Y & 5\\
   \hline   
   \new{V9} & P & F & 24 to 30 & T & 5\\
   \hline
   \new{V10} & P & M & 24 to 30 & T & 4\\
   \hline
   \new{V11} & S & M & 24 to 30 & Y & 6\\
   \hline
   \new{V12} & S & M & 24 to 30 & Y & 5\\
   \hline
 \end{tabular}
 \caption{Participants' background information. For the Participant column, S1-S14 represent the 14 streamers, and V1-V12 represent the 12 viewers. For the Occupation column, P = professional developer (paid to develop software), H = hobbyist programmer, and S = student programmer. \new{The gender information is self-identified by the participants, M = male, and F = female} For the Platform column, Y = YouTube and T = Twitch.}
 \label{participant}
\end{table}

\subsubsection{Interviewee Backgrounds}
Table~\ref{participant} lists part of the participants' background information.
We did not report the streaming content for each participant, as the small size of certain communities might reveal personally identifiable information.
We strove for diversity across multiple dimensions, such as gender, age, and programming languages used in the live streams that participants hosted or watched.
The 26 participants had two main roles: streamers (14), whose primary job is narrating their thoughts along with their programming activities, and viewers \new{(12)}, who watch the stream and interact with others in the same stream via a single thread chatroom. 
For each role, we also required certain activities as part of the participant's eligibility to participate in the study: 1) for both streamers and viewers, the participation frequency of programming-related streams should be at least once a month over the last half a year, so that they were likely to be familiar with certain context-appropriate behaviors, such as viewers responding to one another or streamers engaging viewers in the stream chat, and they should also have previously produced or watched video recordings about computer programming; 
2) for streamers, their streams' purpose should be about sharing programming knowledge and practical skills; 
3) for viewers, they must often watch live streaming sessions for more than half an hour and had some interactions with streamers or other viewers, avoiding those with little engagement. 
We will use S1-S14 and \new{V1-V12} to represent the 14 streamers (S) and 12 viewers (V), respectively. 

Both the streamers and viewers participate in a diverse set of live streaming content, including a variety of programming languages such as JavaScript (8), Python (4), C\# (3), Dart (2), Rust (2), and C (1). 
Out of the 14 streamers, 7 of them primarily stream while working through programming tasks related to open source projects. 
The number of live viewers per session ranges from 5 to 500 people on average.
\new{The majority (17) of all interviewees self-identified as male during our interview.
Regarding the other 9 participants, 8 self-identified as female. 
One streamer preferred not to identify their gender; for convenience, we will use ``non-male participants'' (6 are streamers) to identify these 9 participants through out the rest of the paper.}
Note that the two closest related prior works that focused on programming streams~\cite{faas2018watch,alaboudi2019exploratory} had all male participants.


\section{Study Design Limitations}

While we strove to recruit participants from diverse demographic and professional backgrounds, our participant pool and snowball sampling had some limitations: our sample had only \new{three} female viewers, and all but three participants were from the U.S. The rest were from Europe. Future work could recruit from broader demographics and geographic regions to improve the external validity of our findings.

In terms of participants' motivations, our focus was on those driven by the developer community itself, rather than financial rewards (e.g., leading paid MOOCs) or other extrinsic incentives.
Most of our streamers stream their working on real programming projects instead of conducting a lecture-style sharing of knowledge. 
However, we have seen a few other platforms (e.g., Douyu.com) in which the majority of the live streaming programming channels follow a prepared classroom style, a model to which our findings might not apply.
Also, in terms of streaming platforms,  our participants used only YouTube Live and Twitch.tv, so some of our findings and design recommendations might not speak to those who participate in live streams on other platforms that might have different norms or technical affordances than those we studied.
Future work can compare the norms across platforms by recruiting those who have had related experiences.

\section{Results}
In our interviews with streamers and viewers, we solicited both groups' opinions about their reasons for creating or participating in programming-focused live streams, as well as any barriers they faced with this type of knowledge-sharing medium. In the rest of this section, we present our findings in detail to address our four research questions.

\subsection{Why do streamers choose to share programming knowledge via live streaming?}


We found three main reasons for which streamers would share their programming knowledge via live streams as opposed to recorded videos: 1) video recording takes much longer to prepare compared to live streaming, 2) streaming allows streamers to connect with and update their communities, 3) streamers can receive rewards and assistance from their viewers. We describe each of these reasons in detail below.

\subsubsection{Producing a video takes much longer than live streaming}

When asked why they chose live streaming over video recording, half of the interviewed streamers (7/14) expressed concerns that video recording was a process that could \textit{``tak[e] forever''} (S6), was \textit{``tedious''} (S3), was a \textit{``nightmare''} (S11), and was typically \textit{``really hard''} (S10). 
The dread of arduous behind-the-scenes work along with the slower pace of content generation associated with pre-recorded videos were barriers that the streamers were able to avoid with live streaming.


\begin{quote}
After I've tried live stream [\textit{sic}], I've actually shifted from doing pre-recorded videos because it takes a lot of time to, you know, think everything through, prepare this crap, and record and edit. Then you know, make everything look nice and then publish. While on the live stream, we can just hit the play button and then uploaded [\textit{sic}] to YouTube and you're basically done and people are just as happy about it. -S5
\end{quote}


\new{We also asked streamers why they believed more behind-the-scenes work was needed when recording a video; streamers mentioned that their mindset and behaviors were different between the two activities.} When recording a video, a third of the streamers would shift to a \textit{``flight attendant''} (S9) persona, where the voice-over primarily focuses on the content itself, which is fully prepared and scripted, as if they were \textit{``acting''} for the \textit{``virtual audience''} (S11). In contrast, when live streaming these streamers found their dialogue to be  \textit{``a bit more casual''} (S11, S9) and \textit{``spontaneous''} (S10), and they \textit{``don't care about the risk [of making mistakes] at all''} (S9). 

\begin{quote}
    In prepared content, it's going to be much more scripted. It's going to be much more acted, and I'm going to focus more on my facial expressions and make sure that you know what I intended to be and whatever work I'm creating. I think the prepared content is going to be like an actual performance that you can reshoot and whatnot. But in streaming, you can't really do that unless you're really good at, like, improv acting or something. -S11
\end{quote}

\new{Our finding that streamers adopt different mindsets of preparation for different media is consistent with prior work~\cite{fraser2019sharing} and also similar to viewers' mindsets. 
As we found that before choosing a medium, a viewer has already set certain expectations: for live streaming, they look for explanations and thought processes as well as opportunities for social interaction and casual information acquisition, whereas for pre-recorded videos they seek more specific information and thus expect to see more polished, compact, and edited content. This expectation of casual learning makes the streamers feel less nervous and behave more naturally, resulting in content that is easier to prepare.}


\subsubsection{Streaming is a new way to reach out and update the community}
Half of the streamers (7/14) reported that they often streamed about programming work related to open source projects or online programming communities. 
For example, they would review pull requests, add new features, or resolve repository issues. 
Unlike the other half of the streamers, who have focused on building their own communities with independent projects, these streamers are core contributors in existing open source projects or programming communities. Live streaming allows them to more easily connect with their respective developer communities as a whole, sharing information and updates without excessive effort.

\begin{quote}
    [Live streaming] seems like an opportunity to reach developer communities in a new way that I hadn't tried to access before. -S10
\end{quote}

We also found that streamers use live streaming as a communication platform to share information with the developer communities.
One streamer found streaming to be an effective method of reaching out to viewers in an existing developer community with quick updates on new features.

\begin{quote}
    Two years ago when I started at [Anonymous Community], we had four videos on the YouTube channel, and four of them were from when [Anonymous Community] was, like, an energy startup. So [the videos were] clearly not, like, aligned with our current vision, and having more technical content that I could produce without spending a ton of time on it was a nice way to go back [and update the users]. -S9
\end{quote}


\subsubsection{Streamers receive appreciations and assistance from viewers}

When asked what made them continue streaming, the interviewed streamers mentioned that some of their motivation came from their audience. 
Specifically, the motivators included live interactions around common interests (7), direct assistance on the streaming content from viewers (8), e.g., viewers pointing out the streamer's coding mistakes, and viewers' words of appreciation (6). 
S2 said, \textit{``[for] that pre-recorded material, you're never gonna get feedback on [it], you're never gonna interact with viewers on [it]. It's reference material.''}
Additionally, three streamers benefited from their audience by using them as a sounding board or \textit{``rubber duck''}\footnote{\url{https://tinyurl.com/y333u324}} (S3) for debugging or getting unstuck. 
Furthermore, hearing viewers' comments about how the streams helped them felt personally rewarding for the streamers.

\begin{quote}
    It's just nice to hear when people show up and say `Hey, just wanna let you know I got my first dev job, and thanks so much for keeping your streams on. -S5
\end{quote}

\new{As noted in prior work~\cite{faas2018watch,samat2019live}, the relationship between streamers and viewers on these streaming platforms is unique. Unlike passively listening to pre-recorded information or an authority figure (e.g., teacher), streaming gives viewers the opportunity to shape content through real-time interactions.
We believe that this unique one-to-many model of collaboration helps streamers and viewers bond more closely due to the ease of interaction and the more casual power dynamics of the group.
}

\subsection{Why do viewers choose to learn programming via live streaming?}


Viewers' motivations found in prior work are limited, as some of them are inferred from behavioral observation by researchers~\cite{faas2018watch}. 
Our findings revealed why, when, and how viewers choose whether or not to attend a live stream session, watch a pre-recorded video, or watch an archived live stream. 
We describe each of these reasons in detail below.

\subsubsection{Streaming creates an over-the-shoulder learning experience}


\new{When asked about their goals when watching live streams}, \new{half of the viewers (6/12)} reported that they had no specific \textit{``problems to solve''} (V4,V3) or concrete learning objectives (V1,V5).
\new{Instead, their mindset was to serendipitously learn \textit{``whatever came up that's new for me''} (V8) or \textit{``a better way of what I was doing''} (V9). Such learning moments could also result from the actions of other viewers, who might ask an insightful question or provide new perspectives on a particular problem.}
In contrast, when watching a pre-recorded video tutorial or attending online classes, these viewers mentioned that they often looked for solutions to specific problems or had defined learning objectives in mind. 
Viewers cited a variety of reasons for this, including not approaching a live stream in the same way they would approach a class, not having much information about the live stream provided ahead of time, and expecting to ask \textit{in-situ} questions as the stream progresses. 
In fact, this phenomenon is similar to the concept of \textit{informal learning}, or ``over-the-shoulder learning.'' We will later discuss how live streams can benefit from this practice when used as a form of online learning.

\begin{quote}
    For more specific types of problems that I would have, like...I can also look [them] up on Stack Overflow. Like `how do I fix this error?' So those are not the things that I would watch live streaming for. I'm going to the live stream more for, like, the structure of a project of the [chat]bot. -V4
\end{quote}

\new{When asked about their overall learning experiences}, most viewers reported satisfaction with the streams that they have watched fully, although only a few of them directly applied the techniques learned from the stream to their daily work. 
\new{They said most learning happens when streamers share tips and tricks (V3, V11) or real-life uses for the streamed techniques (V9), or when streamers explain the underlying structures of problems or applications (V4, V7).}

\subsubsection{Viewers like to see and hear streamers' thought processes}

\new{When asked about their expectations of the streamers, most viewers (10/12) commented that they like to see streamers share the thought processes behind the structure of their code, how they debug issues they encounter during the stream, and how they learn new methods or techniques as they work.}
This finding is consistent with prior work on the motivation of viewers watching live programming~\cite{raj2018role,faas2018watch}.
As viewers often bring this expectation to live streaming sessions, they may find it unsatisfactory when streamers are not actually live streaming but instead \textit{``copying and pasting''} (V6) pre-written code. 

\begin{quote}
    I like to see how the person thinks and makes mistakes, and a lot of people edit it (the video) too harshly. And then it ends up being this one fluid front-to-back program with no mistakes. They are not researching anything or looking anything up, and that's not realistic to me. So I think it's a little bit more entertaining, a bit more easy to sit through a lot of the live stream videos or even their archive videos. -V4
\end{quote}

\new{Nonetheless, two streamers (S1, S4) worried about not being prepared enough. These worries led them to either avoid talking about certain materials to avoid potentially getting stuck on bugs (S4), or even not live streaming at all in some cases.
We discuss similar trade-offs and misaligned expectations in our discussion session and design recommendations.}

\subsubsection{Real-time interaction makes the learning more authentic}

\new{When asked how live streaming experiences differ from video recordings, viewers reported feeling more authentic, realistic, and reliable (V1, V2, V4, V10) when participating live.}
V2 said \textit{``It feels better, feels [like] you can interact with the actual streamer live and ask questions.''} 
The real-time proceedings make viewers feel that \textit{``[streamers] can answer questions from you right there''} (V1).
\new{However, viewers did find video recordings to be more complete and of higher quality, while live streams were often considered too long (V9) and \textit{``a bit junky''} (V10).}

Half of the viewers (6/12) seek closer connections with others in the community as opposed to a one-off encounter. \textit{``I was looking for communities that I can join in, and [the streamers] are almost like a mentor.''} V4 said.
Similar to prior work on the live streaming of other content~\cite{fraser2019sharing}, viewers also expressed their preference for watching certain live streams, as they favor the way that the streamer explained things (V1, V3, V4, V5, V7, V11).



\subsection{What are the barriers that streamers face when sharing programming knowledge via live streaming?}

In this section, we report difficulties streamers face when creating content for programming-focused live streams.
We address topics that have not been considered in prior work, including privacy, preparatory work for a streaming session, and issues that members of marginalized groups face.



\subsubsection{\new{Preventing privacy leaks is a demanding and tedious task}}

\new{Privacy leaks are one of the biggest concerns the streamers had. 
Admittedly, streaming platforms have already taken actions to clear their chats of discouraging content, including targeted harassment, the unauthorized sharing of private information, hate speech, and spamming, among other things~\cite{privacynews1,privacynews2}.
However, the nature of live streaming and the ability for anyone to join a stream means that streamers must constantly consider their privacy practices.}
During our interview, all streamers had concerns regarding the potential risk of exposing their personal information to an unknown audience. This information ranged from digital information such as email addresses, passwords, and API keys to physical information such as \textit{``where I live''} (S5) or \textit{``my family''} (S2). S2 raised some concerns: 

\begin{quote}
    [It is] very easy to dox yourself. I have had guests on who inadvertently showed their phone number on stream. I've had issues where the API keys for various services have inadvertently been shown on stream.
\end{quote}

The sensitivity of information could also depend on the content being streamed. 
Three streamers had live-streamed using products from the organizations they were work for, and all of them expressed the challenge of keeping their organizations' information confidential, including \textit{``unannounced projects''} (S11) or \textit{``user-related information''} (S9). S9 said, 

\begin{quote}
    My number one thing is that I will I never, ever, ever want to violate user privacy. That's why I never use a dev environment. I never use my admin account. I actually have a second non-admin account that I use [for streaming].
\end{quote}


\new{We followed up on these concerns and asked streamers about their current practices to prevent privacy leaks.}
All streamers explained that they prepare to live stream by performing a series of protective actions to safeguard their sensitive information. 
Before live streaming, streamers log in to all necessary accounts (14 streamers), use anonymous mode in their browser (10), share only part of the screen (8), and use unrecognizable usernames or accounts only while streaming (6). Some streamers also consider the background of their videos as well:

\begin{quote}
    [The] easiest way is just to keep it away from anything related to the street, so I don't have any windows behind me. It's hard for people to tell where I am. I'm just doing it in an all closed room right now -S10
\end{quote}



Nearly all streamers (12/14) expressed the tedium and effort involved in these preparation tasks, and even if these streamers are thoroughly prepared, we found that they were still worried about potentially leaking their private information during stream.
These streamers noted that they must carefully monitor both their current shared content as well as what they are about to share and say to avoid leaks. S3 said, \textit{``recently I do work on [\textit{sic}] projects in private repositories that aren't public. I'll need to keep in mind not to just open GitHub in the front page on the live stream. So I need to just go to a repository for example.''} 
Even with such efforts, these streamers still occasionally encounter unintended information leaks: 

\begin{quote}
    You code on screen, you dump an object [that] has a bunch of secrets in it. It happens. I average like one secret getting leaked out like every six hours of screen time. -S10
\end{quote}

\begin{quote}
    You don't really realize how many times like a random API key or some other secret is in log files until you try and code on the stream. -S10
\end{quote}

When we asked those streamers who had privacy concerns or incidents how they prevent or handle these incidents, the answers were mixed. Half of the streamers (6/12) reported that they do not do anything, as they think \textit{``nobody made a big deal out of it''} (S9), and they trust their technology and platform to protect them (S10). 
The rest were very sensitive and cautious about their privacy and often applied advanced precautionary practices, such as \textit{``[muting] my mic while typing my password''} (S7), or \textit{``[notifying] the SWAT team in my city''} (S10). 
\new{These mixed strategies suggest that more privacy protection guidance is needed on the streamer side of streaming platforms.}

\subsubsection{Gender-related harassment is common and difficult to prevent}

\new{Prior work that studied live streaming programming had all male participants~\cite{faas2018watch,samat2019live}; nearly half of our streamer participants (6/14) self-identified as women, and one preferred not to say. 
When asked about gender-related issues,}
more than half of these non-male streamers (4/6) said their biggest concern is gender harassment, such as receiving comments about their physical appearance. 
In contrast, none of the eight male streamers mentioned gender harassment as a concern during their interviews. 
One non-male streamer noted the frequent occurrence of these unwanted comments: 

\begin{quote}
    For me, it's all gender aspect. So [people just keep] making comments about irrelevant things like my physical appearance and things like that. -S10
\end{quote}

These comments are often from the live chat, which is difficult to ignore or filter out. 
However, some comments are made outside of live stream: \textit{``I tend to get more of the creepy messages to my email or to my Twitter''} (S12). 
These comments have made streamers feel \textit{``horrible''} (S9), \textit{``uncomfortable''} (S12), and worried about their safety (S10). 
\new{Although this harassment issue echoes research in other streaming contexts~\cite{wohn2019volunteer,seering2017shaping}, all non-male streamers felt that the developer communities are \textit{``pretty respectful''} (S12) and \textit{``open to LGBT communities and friendlier in general''} (S11).}

\new{Much prior work has shown that the field of software development remains dominated by men~\cite{imtiaz2019investigating}. As we reported, the synchronicity of live streaming brings more uncertainty, making it harder to control for streamers. This might further discourage non-male programmers from engaging in the programming community.}
When asked why there are few non-male programming streamers, all six non-male streamers cited the imbalanced gender ratio in developer communities and the hassle of dealing with potential harassment as two main reasons. 
S11 said, \textit{``there are challenges that men just don't face when doing this type of stuff online. Men generally don't have or generally aren't being harassed for their gender.''} 

\subsubsection{Challenges and practices that the non-male streamers face when streaming}

\new{We followed up with these streamers about their current practices and advice for handling harassment.}
None of the streamers believe the current tools on their streaming platforms are sufficient to prevent harassment. \textit{``I wish I had one better set up so I could like block people from within the overlay''} (S12). 
When harassment occurs, streamers often warn and ban the offending commenters immediately while working to build a healthier community \news{ --- an effective strategy according to prior survey results~\cite{cai2019effective}}.

\begin{quote}
    If there are people who troll you [...] you can get rid of them as soon as possible, then you're going to start building a community that's really supportive of you, and that support gives you a lot of strength. -S11
\end{quote}

Besides tools, three of the six non-male streamers mentioned that finding other streamers who share a similar identity has helped them overcome this barrier: \textit{``one of the reasons why I felt like it was OK to start was I saw another person [like me] doing it''} (S12). 
We also found that non-male streamers spent more effort signalling their ability during the interview; one noted that \textit{``I think it also helps that I have a PhD''} (S9).
All six non-male streamers talked about their qualifications in terms of external validations (e.g., a certificate or a degree), while none of the male streamers we interviewed mentioned such.

\begin{quote}
    [I] wrote the standard introduction book to [framework], and so most of the people in the [framework] world---not all, but most of them---are pretty respectful. -S12
\end{quote}

\subsection{What are the barriers that viewers face when learning programming via live streaming?}


\news{Viewers also reported multiple barriers to learning via live streams. Beyond the low participation rates reported in prior work, we also found that viewers had trouble efficiently finding personalized learning resources, such as appropriate streams to watch or resources mentioned during a stream session. 
Viewers often had to resort to impromptu research (e.g., a Web search) during the stream and try not to fall behind, which can create intense time pressure.
}

\subsubsection{Related resources are not shared in real-time}

\new{When asked about counteracting challenges experienced during streams, all viewers wished to follow streams without worrying about missing useful content while they researched new information and questions on the side.}
V4 wished for the code and resources to be shared \textit{``on the fly.''}
\new{Three viewers wished to receive basic information about the stream, such as prerequisite knowledge, as well as related materials, such as starting code, before the beginning of the stream.}

\begin{quote}
    It would be good for the streamers, maybe before they do the live stream, [to] make an announcement and they release, like, resources that they're going to use during the coding. [...] Other people can definitely follow along when they have a package and stuff. - V1
\end{quote}

However, the live nature of streaming means that streamers often improvise their knowledge sharing. This modality makes it challenging for streamers to share materials in advance, since live streams are not typically scripted or planned concretely. 

\subsubsection{Finding high-quality and helpful live streams can be difficult}


\new{Choosing what to watch can be a non-trivial process for some viewers. Prior work found an effective streaming title is important to attract viewers~\cite{tang2016meerkat}.}
\new{Similarly, when asked how they discover and decide to watch certain live programming streams,} some viewers reported behaving as``shoppers'' looking for the right match, but this can be an inefficient process. ``I [don't] realize this topic isn't that much of interest to me till the  streamers [start] coding, which is 20 minutes in already'' (V5).
\new{Other viewers chose to only attend streams hosted by streamers they were already familiar with. This might limit their opportunities for learning from other channels. Fundamentally, this is because} unlike recorded videos, which include indicators of video quality (e.g., number of views, likes, comments) and the ability to scrubbing through the timeline, there are often limited signals to determine the quality of a live stream. 
\new{For example, V8 said, ``I have to listen to the agenda at the beginning to make the decision.''
When joining in the middle of the live stream session, viewers often have to spend some time listening to the content to see if it is the right fit (V2, V10), or rewinding the video to find indications of video quality and relevance (V12).
}

Additionally, streamers usually do not share their streaming resources (e.g., code, documents)  in advance, so it is difficult for viewers to quickly search and find a stream they would like to watch. 
Although the viewers in our study were experienced with live streaming, some of them still made their decisions solely based on the title of the stream, which could be ineffective.

\begin{quote}
    I just look through and go by what the title is describing. And then after, I start watching it for like maybe 5 - 10 minutes. If it's going too slow or [it's] just not [about] topics that I'm interested in, then I'll probably switch and find something [else]. -V5
\end{quote}

Other viewers found their own methods of choosing a stream to watch using indicators such as their prior experience with a particular streamer or examining the streamer's video archive. 
For example, V4 relied on their knowledge about the streamer's credentials to assess content quality: \textit{``[the streamer] has a PhD in computer science, and so when I watch it, I know that I'm getting legitimate, helpful knowledge.''}




\subsubsection{Impromptu research may be needed when encountering new concepts}

\new{Because viewers sometimes watch streams without specific questions or learning objectives in mind, as we reported above, they may unexpectedly encounter new concepts during live streams and need to engage in impromptu research to learn more. }

\begin{quote}
    I would listen to the live stream, and then as they were talking I'd be, like, googling things or googling more about certain topics. -V5
\end{quote}

As mentioned earlier, this style of learning is similar to ``over-the-shoulder learning,'' in which a learner looks over someone else's shoulder as they work on a project on screen~\cite{bannon1986helping}.
However, there are always trade-offs between time spent watching a stream and looking up newly encountered concepts that are useful to learn.



\subsubsection{Learning experience is not personalized to the viewer's ability and preferences}

\new{Viewers expressed a desire for easy access to reference materials as the stream plays, as \textit{``it would be cool to see some sort of live description box [on the side]''} (V2).}
\new{Other viewers suggested techniques to help address the misalignment between their learning pace and streamers' teaching speed.} The fact that live streaming is free and has no minimum requirement for the viewer's level of expertise also means that viewers often have more diverse backgrounds than those who attend structured classes. 


We also noticed that different learning paces influenced viewing behaviors.
\new{We found that one viewer coded along with the streamer consistently (V6); four took notes (V1, V4, V6, V12); seven watched the stream while taking notes, coding, or intermittently researching concepts (V4, V5, V7, V8, V9, V11, V12); and three watched the stream without engaging in other behaviors (V2, V3, V10). }
However, because of the ``live" nature of streaming, these viewers often felt they had \textit{``fallen behind''} (V6) or were \textit{``lost''} (V4) but did not want to interrupt the streamer due to perceived social norms. 

\begin{quote}
    [I've] fallen behind because I'm doing research or just because he's moved fast. It would be pretty awesome [to] sync with him again, like [I'm] desperate [\textit{sic}] at that point because I can't catch up to him if there's something wrong or I've missed a line or a typo and he's just cranking on.
\end{quote}

\new{Unfortunately, struggling to strike the right balance between getting hands-on experience and following streamers closely might make viewers miss certain learning opportunities. }


\subsubsection{\new{Chat can be another place to learn, yet it often lacks context for others to understand}}

When asked about the live Q\&A experience with viewers, all streamers reported that for most of the time they have no problem understanding and responding to viewers' questions in the chat. 
This makes sense, as viewers' questions are often related to concepts that streamers have recently discussed.
\new{Meanwhile, fewer than half of the viewer participants (5/12) found the chat occasionally useful: ``other people might sometimes have a similar question like you, and you got to know more [about the] thought process of solving one problem from [other] people's perspective'' (V9). At other times, viewers either hid the chat window (V1-3, V5, V6, V11) or didn't pay attention (V4, V7, V8, V10, V12), and some wished to have a filter to remove irrelevant chat messages (V7,V9).}

\new{We found that both streamers and viewers had the experience of not being able to understand others' conversations in the chat.}
The challenge of understanding others' chat messages arises from the loss of a shared context and the large volume of chat messages, which has been mentioned in related work~\cite{lu2018streamwiki, yang2020snapstream}. 
In the context of programming, we observed the challenge comes from the fact that the code is constantly changing, which means viewers' references (e.g., line number) are only correct at the time the information appeared in the chat. 





\subsubsection{Stream archives can enable deeper use of the content but are hard to navigate}

\new{Consistent with prior work~\cite{fraser2020temporal,lu2018streamwiki}, watching stream archives is also a common practice among programming stream viewers.}
The majority of our viewer participants (9/12) mentioned that the information they seek when watching live stream archives is different from the information found in pre-edited videos. 
This information includes a streamer's thought process (V3-6), Q\&As (V1, V7, V12), recently-introduced techniques and features (V11, V12), or the social aspects of a stream (V9).
This aligns with our aforementioned findings regarding viewers' motivations: viewers do not often decide to watch live streaming to solve a specific problem. Instead, they are typically patient, relaxed, and open to receiving new ideas when participating in live streaming or viewing archives. 

Additionally, six viewers have revisited stream archives after attending the live version.
\new{This often occurs when viewers consider certain parts of the stream \textit{``interesting''} (V4, V5), need to re-learn specific techniques or knowledge (V7, V12), or encounter complex new content that they could not digest well while participating in real time (V6, V11).}
Watching stream archives allows them to further analyze this knowledge and learn at their own pace.

\begin{quote}
     The topic is just hitting me the first time, and then if I notice later [that] I was like `oh, I really have to get into this and use this,' that's, like, the first wave of knowledge and information about this thing. Then I'll go back and be like, let me break down this live stream or whatever I watched and, you know, plan out some topics that I'm going to go over myself and get more deeply into it and figure out what's up. -V5
\end{quote}

When asked how they navigate the stream archives, seven viewers mentioned that they often found it difficult to locate specific information that they were looking for. 
\new{Instead, they would speed the video up or scrub the play bar to retrieve content that seemed relevant, which was inefficient and might lead viewers to miss important information (V7).}
This is because videos are often unplanned, and there are no timetables to facilitate skipping to specific content. Consistent with prior work~\cite{fraser2020temporal}, live streams are often very long, so useful information is more likely to be scattered throughout the video, and since the stream is not edited, not all of the content may be engaging or helpful. 


\begin{table}[b]
 \begin{tabular}{ p{9cm}| p{4cm}}
  \textbf{Main Results}	& \textbf{Design Implications} \\
  \midrule
  \parbox{9cm}{\vspace{0.3pc} - Related resources aren't shared in advance \\
- Finding quality, helpful live streams can be difficult\\
- Impromptu research may be needed if viewers watch streams without specific problems in mind\\
- Certain techniques help create a more personalized learning experience \vspace{0.3pc}} & \parbox{4cm}{\vspace{-4pc}Designing personalized live learning environments} \\

   \hline
   \parbox{9cm}{\vspace{0.3pc}- Chat rooms lack methods of capturing context \\
   - Stream archive is useful but hard to navigate \vspace{0.3pc}} &  \parbox{4cm}{\vspace{-0.1pc}Creating new tools for content retrieval and understanding} \\
    \hline
    \parbox{9cm}{\vspace{0.3pc}- Strategies to prevent privacy leaks can demanding and tedious to put in place \vspace{0.3pc}} &  \parbox{4cm}{Easing the effort of privacy protection}\\
     \hline
     \parbox{9cm}{- Gender-related harassment is common and hard to prevent} & \parbox{4cm}{\vspace{0.1pc}Designing healthier and cleaner learning environments\vspace{0.1pc}}\\
      \hline
 \end{tabular}
 \caption{Four design implications that are drawn from our results.}
 \label{design}
 \vspace{-1pc}
\end{table}

\section{Discussion}

In light of our findings, how can we redesign streaming platforms to better support this informal learning and teaching practice in developer communities? 
To address this question, we discuss four design implications in this section (Table ~\ref{design}).

\subsection{Designing Personalized Live Learning Environments}

\new{Developing online personalized learning environments (PLEs) has been explored for more than a decade~\cite{attwell2007personal}.  
Prior work has recognized the importance of informal learning beyond the traditional educational system~\cite{dabbagh2012personal}.
Our study findings have shown that live streaming as a new learning practice contains the informal learning characteristic, yet current streaming platforms provide insufficient guidance and support for either streamers or viewers to effectively leverage and benefit from the pedagogical advantages of informal learning and social media.}


\new{As Attwell also advocated, there should be more attention paid to informal learning by extending access to the educational technologies that have only been made available to those formal educational programs~\cite{attwell2007personal}.
From our observation, the participants' behaviors were driven by the streaming platforms, from their user interface (UI) design to the social norms among their users.}
The UI of these live streaming platforms are inherited from the versions designed with only gamers or entertainers in mind, not learners and teachers.
The design principles used on these platforms focus on optimizing the UI for extended visual attention, such that viewers are less likely to look away from streamed content. 
However, both prior work and our research have shown that viewers often need to look away from the video stream in order to digest the incoming knowledge through complementary activities, like asking questions, taking notes, or doing relevant exercises~\cite{national2009learning}.
The behaviors that benefit learners most are not well supported by live stream platforms designed for entertainment content. 
This indicates a need for platform design updates that consider the context of learning, allowing platforms to meet the needs of creators and viewers of educational content.




Before online learning became popular, the education community was already grappling with the challenge of providing appropriate educational opportunities at scale in traditional in-person settings (e.g., in large lectures)~\cite{psotka1988intelligent}.
As we found, the difficulty in adapting teaching and demonstration to viewers is due in part to their differing expertise levels. 
Below, we discuss a few design implications based on our findings and prior work on learning at scale and personalized education.

\textbf{Scale learning with better real-time information acquisition tools.} The viewer's challenge of continuously following along at a streamer's pace also occurs in other groupware settings, such as in virtual meetings~\cite{junuzovic2011did}. The lack of support for latecomers who must obtain missed information could negatively affect their engagement and overall productivity. Prior work proposed catch-up techniques for viewing and replaying missed activities in settings that require real-time interactions among stakeholders~\cite{gutwin2017looking}. 
In the context of live streaming programming where the viewers shift their attention away from the stream and want to sync back up again later, we propose the \textit{pause} feature from prior work~\cite{lasecki2014helping}. 
This allows viewers to pause, fast-forward, and jump immediately back to the current point of the live stream. 
Furthermore, when there are multiple files and applications that the streamers were switching in between, this feature would allow viewers to stay at the view (e.g., a code file) that they want to pay more attention on.


Streaming platforms could also integrate information seeking tools to ease the process of knowledge acquisition in real time.
We propose that future streaming systems could automatically provide related resources as streamers narrate, such as specific programming concepts or sections in the documentation. 
This could be done by combining the techniques of auto-transcription and Google search APIs. 
Along with this feature, we suggest that platforms enable real-time code sharing, with streamers' permission. 
This would help those viewers who code along but worry that they ``would have fallen behind'' because they took the time to look up a new concept. 
Prior work has proposed effective tools with automation techniques~\cite{head2015tutorons, brandt2009two,liu2019unakite}, or crowdsourcing approaches~\cite{chen2017codeon,latoza2014microtask} to provide easy access to code-related support. 
Future work could explore more techniques to help viewers to learn at their own pace.

\textbf{Personalized stream matching process.} 
\new{Our finding suggests that due to the lack of content selection signals, viewers put more weight on the limited information available to make a decision about what to watch. 
This selection strategy resonates with findings from prior work on the large amount of effort that streamers put on branding their channels~\cite{pellicone2017game}. }
In the future, we can also consider mechanisms that help align a stream's difficulty level with the viewer's current knowledge and skills. 
This will help move towards a constructivist learning environment where viewers are more likely to retain knowledge attained by engaging in contextualised problem-solving~\cite{pears2007survey}.
Examples could include showing a visual ``difficulty'' badge for an upcoming stream, including a section of knowledge prerequisites in the stream's description and real-time displays of viewers' self-reported expertise level. 
This could allow streamers to adjust their teaching strategy and content, shaped indirectly by viewers, and potentially help indecisive viewers determine whether they should join the stream. 
Future work would be needed to conduct deeper analysis on streamers' live streaming styles to match them with viewers' viewing preferences. 



\subsection{Creating New Tools for Content Retrieval and Understanding}
A major challenge for many of our viewers was the difficulty of retrieving and understanding information that they missed or revisited. 
This information was mostly from the previous chat and the video content.
In particular, the lack of context in the chat and poor tool support on video navigation can make learning and communicating with each other more difficult.
We envision two directions for designing tools to facilitate retrieving and understanding information: context-captured chat interfaces and lightweight stream archiving tools.

\textbf{Context captured chat interface.}
The current chat tool on streaming platforms does not support context referencing (e.g., line number), which makes the chat hard to understand during programming streams.
Additionally, the fact that streamers constantly change their code makes it more challenging to understand past references in the chat.
\new{To address this issue, Snapstream~\cite{yang2020snapstream} allows viewers to snapshot the video content as shared context to ground their chat.} 
Chat.codes~\cite{oney2018creating} and EdCode~\cite{chen2020edcode} enable students and teachers to link code snippets to their Q\&A messages for context reference.
However, our problem requires context that might link to multiple points of the video. 
Similar to prior work that automatically segmenting live stream archives~\cite{fraser2020temporal}, we suggest developing a chat interface that is automatically organized by anchor points, \news{such as short clips in the streaming content or keywords in the auto-generated transcript}.
In this way, the system would automatically anchor and manage the stream content based on the chat time and natural language context, \news{making it easier to navigate.}








\textbf{Lightweight stream archiving tools.} 
Our findings have shown insufficient support for interacting with stream archives efficiently. 
This includes 1) the presence of uncut irrelevant content (e.g., idle stream time), 2) no navigation support (e.g., a timetable), and 3) a lack of relevant and interesting interactions (e.g., q\&a in the chat) during the stream.
All streamers we interviewed had a desire to better archive their streams for various benefits, such as to build viewership or for future content reference. Some of them expressed their willingness to add timetables, but they often felt ``tired'' after a few hours of live streaming, making it more difficult to follow through.



To address this challenge, prior work has attempted to leverage viewers to help with annotation and documentation of the stream archive~\cite{lu2018streamwiki}. 
However, they found that viewers often had low incentives to contribute to the work, which we found from our interviews as well.
To achieve a similar goal, \new{Fraser et al. used automated tools to segment live streaming artists' videos~\cite{fraser2020temporal}. However, their algorithm requires application usage logs, which can be challenging in programming streams due to the obligation to constantly switch between applications.}
We suggest designing a lightweight annotation tool that allows streamers or their designated assistants to take notes as they stream along, with automatically captured timestamps and auto-generated transcript. 
Once live streaming is over, the timetable would be automatically generated.

\subsection{Reduce the Effort of Privacy Protection} 

The challenge of protecting privacy is much more difficult in live streaming than it is in pre-recorded video. 
Unlike pre-recorded video, where people can always edit their recordings before posting, the nature of live streams makes it infeasible to edit their recording~\cite{lee2019effect}.
Writing scripts for the stream would not work, as interacting with the live audience would break the plan by introducing new content, which could result in streamers showing or saying sensitive information. 
Streaming programming work makes privacy control more challenging, as streamers constantly switch between applications (e.g., IDE, browser, terminal) for different coding-related tasks.

\new{Platforms such as Discord~\cite{discord} have released features to help protect streamers from accidentally revealing sensitive information. However, these features are rule-based (e.g., hide email) and do not effectively scale to the improvised situations.}
Prior work has studied people's privacy concerns regarding information disclosure during others' live video game streaming~\cite{li2018tell}. 
Given that their main findings are concerns a participant might have regarding the team she/he was in, their design suggestions do not apply in our case.

One approach would be to simply provide a privacy checklist that guides streamers through the steps that they need to take before streaming.
An automatic checker could also require that certain privacy settings are in place, such as sharing a browser window in incognito mode.
Additionally, integrating with existing automated tools, such as OCR, can help detect patterns that are potentially sensitive, such as cell phone numbers, emails, or home addresses, and then we can use the existing user interface design to hide them or notify streamers.
The computational cost of conducting OCR can be high when done in real time. 
Instead, a post-hoc video process would be better to help reduce the risk of exposing sensitive information to malicious viewers through the archived stream. 



\subsection{Designing Healthier Learning Environments}



While enhancing streaming techniques can help address certain problems in programming education, the solution to many other issues in this emerging sociocultural environment may exist fully or partially outside the purview of tools and technologies.
Gender-related harassment, in particular, has been an issue in live streaming in other content areas, such as gaming~\cite{wohn2019volunteer}.
In developer communities, gender-related harassment in live streaming is not yet well-studied, but the general gender imbalance likely makes the issue prevalent. 
Prior work has discussed such issues in the developer community on sites such as GitHub~\cite{imtiaz2019investigating}.
Aside from gender-related harassment, we also observed skill-related stereotypes from the audience. To address the problems we found, we suggest a few social strategies and design adjustments to benefit both platforms and streamers wishing to create a healthy and safe learning environment. 

\textbf{Promoting underrepresented streamers.} 
\new{Streaming platforms should help signal their ability with tags so that streamers do not have to always mention it. Also, platforms can promote these streamers' channels even to other relevant stream topics (e.g., drawing) to attract people with similar identity.}

\textbf{Improving usability of the moderation tools.} 
A few streamers mentioned that they did not set up the platform's built-in moderation tool because they perceived it to be too complicated to initiate. 
Streaming platforms could make the moderation setup process easier in the future. 


\textbf{Incentivizing viewers to help moderate the chat.} Prior work has studied moderation in live streaming for games. Researchers found that viewers' motivation to volunteer to help moderate streams is not very high. During our research, we asked viewers how willing they would be to help streamers to moderate the chat. Some viewers expressed concerns, such as the excessive time and commitment needed, and noted that their motivation depends on the streamer in question. For viewers who have followed a certain streamer for an extended time or who have developed a close relationship with the streamer, they expressed a willingness to help moderate the streamer's chat.  In the live streaming programming community, the audience size is usually not large, making it even more difficult to find a volunteer to help moderate for the entire session. We propose that the streamers or the streaming platforms could develop incentives, such as virtual points and badges or privileged access, for those who moderate the chat. They might also allow viewers to volunteer for only a portion of the entire streaming time.

\section{Conclusion}

In this research, we sought to understand the benefits and challenges of live streaming as a new way to share and acquire technical knowledge. In particular, we aimed to position live streaming in the ecosystem of online programming education by comparing it to traditional video-based learning experiences. Through in-depth interviews with 14 streamers and 12 viewers who regularly participate in programming-focused live streams, we identified the unique educational value live streaming provides due to its improvisational, real-time, and social nature. We found that these characteristics of live streaming not only lower the barrier to sharing knowledge but also allow the viewer to shape the content being taught while observing error recovery and workarounds they would otherwise miss in well-edited videos. Live streaming thus plays an important gap-filling role in the rather crowded space of online learning technologies.     

Nonetheless, our study also revealed a number of challenges users face in education-focused live streaming. 
In addition to a lack of features specifically designed to support learning in video live streaming platforms, streamers also have to deal with privacy risks and sometimes online harassment. 
Drawing from our findings, we suggest better tools and strategies for leveraging the benefits of live streaming and mitigating the drawbacks and risks of participation. 
We contribute a critical examination of this young practice to the HCI literature and several design directions to evolve live streaming platforms into safer, more effective, and more inclusive learning environments.

\section{Acknowledgement}
We thank our participants for sharing their experiences and perspectives with us, Google for sponsoring this work, Rebecca Krosnick for her feedback and proofreading the writing, and all of our reviewers for their constructive feedback on this paper. 

\bibliographystyle{ACM-Reference-Format}
\bibliography{sample}

\end{document}